\documentclass[12pt]{article}
\usepackage{amsmath}
\usepackage{color}
\usepackage{graphicx}
\usepackage[footnotesize, skip=-25pt]{caption}

\usepackage{subcaption}

\usepackage{graphicx}
\usepackage{amsmath}%
\usepackage{amsfonts}%
\usepackage{amssymb}%
\usepackage{graphicx}
\usepackage{subfig}
\begin{document}

\begin{center}
{\Large\bf Inflation due to non-minimal coupling of $f(R)$ gravity to a scalar field}
\vspace*{1cm}

{\Large Romy H. S. Budhi}~\footnote[1]{e-mail:
~romyhanang@ugm.ac.id}, ~
\vspace*{0.5cm}\\

$^1${\it Physics Department, Gadjah Mada University, 
Yogyakarta 55281, Indonesia}\\
\end{center}
\vspace*{1.5cm} 

\noindent
{\Large\bf Abstract}\\
In this work we investigate a  inflationary scenario generated by a large scalar field $\phi$ that non-minimally couples to a $f(R)$ modified gravity model. For a Starobinsky's like model,  it is found that along a particular flat direction, the scalar potential takes a simple form $V = (M_p^4/4) [V(\phi)/\alpha(\phi)^2]$ where $\alpha(\phi)$ is a non-minimal coupling to Ricci scalar $R$ in the model. The inflation, therefore,  is effectively represented as a single field inflaton scenario.  For a specific example, such as a scalar potential  $V(\phi) = \mu_1\phi^2 + \mu_2\phi^4$, we found  that the predictions  match nicely in the $1\sigma$ confidence level of \textit{Plank} TT, TE, EE+lowP combination data of Planck 2015 CMB data for $ 0<\mu_3 \leq 100$, where $\mu_3 := |\mu_1|/(\mu_2 M_p^2)$. For example, taking $\mu_3 = 0.01$ the scalar-to-tensor ratio $r=0.0004$ and and spectral index $n_s = 0.96985$ for $N_\ast = 50$ while taking $\mu_3 = 100.0$ produces  $r= 0.03$ and $n_s = 0.96359$ for $N_\ast = 60$. 

\newpage

\section{Introduction}

Inflation, a rapid accelerated expansion in the early universe,  was originally introduced to solve the flatness and horizon problems in the big bang cosmology. The inflation provides a mechanism to generate  density fluctuation which later evolve into large scale structures in the universe. Its prediction to  the almost scale-invariant spectrum of cosmological perturbations is in remarkable agreement with the high-precision CMBR observations such as the COBE, WMAP, PLANCK,and BICEP2.  The recent Planck-2015 mission measures the scalar spectral index as $n_s = 0.968 \pm 0.006$ and combined  with the BICEP2/Keck Array CMB polarization experiments have put a bound on the tensor-to-scalar
ratio as $r < 0.12$ (95$\%$ CL). These constraints imply that a single field inflation models with quartic $\lambda \phi^4$ and quadratic $m^2 \phi^2$ potentials are ruled out from the observations as
they produce large tensor-to-scalar ratio $ r \simeq 0.26$ and
$r \simeq 0.13$, respectively, compared to  the inflationary scenario motivated from modified gravitational sector view point, such as $R^2$ correction known as Starobinsky model that predicts smaller scalar-to-tensor ratio \cite{Planck2015}.  A novel scenario from  particle physics view point that is mathematically equivalent to the $R^2$ inflation is now known as the Higgs inflation scenario \cite{Higgs}, where the inflaton field $\phi$ is non-minimally coupled to the curvature scalar R. However, due to a very large non-minimal coupling to the curvature scalar to obtain the observed CMB amplitude, such a model however encounters the problem of unitarity violation in Higgs-Higgs scattering via graviton exchange at Planck energy scales \cite{Unitary2, Unitary3}.   On the other hand, a generalization of scalar-curvature coupling $\xi \phi^2 R$ of Higgs inflation to $\xi \phi^a R^b$ eliminates this large coupling problem \cite{fR1}. However,  this  kind of generalized non-minimally coupled model with an additional  quantum corrected quartic
potential $\lambda \phi^{4(1+\gamma)}$, which is equivalent to the
inflation model $R + R^\beta$ \cite{Mod}  produces
large $r \simeq 0.2$, and thus is disfavored by present CMB observation \cite{Planck2015}. Inspired from this trend, in this paper,  we would like elaborate non-minimal coupling of a scalar field to the curvature scalar in the context of general modified gravity $f(R)$ theory. To have a comparison with the previous studies, a specific $f(R)$ and scalar potential will be considered in the detail.

\section{Action of $f(R)$ gravity with non-minimal coupling to a scalar}

The action for a scalar field $\phi$ in a $f(R)$ gravity theory  can be written as 
\begin{align}\label{1}
	\mathcal{S} = \int d^4 x \sqrt{-g} \left[ f(R) -  g^{\mu \nu} \partial_\mu \phi \partial_\nu \phi - V(\phi)\right] 
\end{align}
in the \textit{Jordan frame}, where here $g^{\mu \nu}$ denotes metric tensor of the space. If there are non-minimal couplings between  $f(R)$ gravity with the scalar $\phi$, to accommodate such couplings, the function $f(R)$ should include any term containing  Ricci scalar $R$ and $\phi$. Therefore, the action can be modified to be \cite{fR1} :
\begin{align}\label{2}
	\mathcal{S} = \int d^4 x \sqrt{-g} \left[ F(R,\phi) -  g^{\mu \nu} \partial_\mu \phi \partial_\nu \phi - V(\phi)\right]. 
\end{align}
The gravitational part of the action may be assumed to be generated by an auxiliary scalar field  $\chi$ \cite{fR2, fR3, fR4}, so that in this context  $F(R,\phi) = F(\chi, \phi)$. Without losing generality, if we work in  arbitrary $D$ dimension space-time, the associated gravitational action term satisfies
\begin{align}\label{3}
	\int d^D x \sqrt{-g} \left[ F(R,\phi) \right] = \int d^D x \sqrt{-g} \left[ F_\chi (R-\chi) + F \right],
\end{align} 
where $F_\chi:= \partial F(\chi, \psi)/\partial\chi$. Varying the action (\ref{3}) with respect to  $\chi$ yields conditions $F_{\chi\chi} = 0$ atau $\chi=R$, showing that it is classically equivalent to the original model if we take  $\chi=R$.  Thus, rearrangement of equation (\ref{1})  leads into this following form
\begin{align}\label{4}
	\mathcal{S} = \int d^D x \sqrt{-g} \left[ F_\chi R  -  g^{\mu \nu} \partial_\mu \phi \partial_\nu \phi - \left(V(\phi) - (F - \chi F_\chi)\right)\right].
\end{align}
 
The action (\ref{4}) can be rewritten to the Einstein frame by conformal transformation \cite{Conf}
\begin{align}
    \hat{g}_{\mu\nu} = \Omega^2 g_{\mu\nu},
\end{align}
to gives rise to the standard Einstein-Hilbert action containing $(M_p^2 R/2)$ term where $M_p$ denotes (reduced) Planck mass in $D$ dimension which is $M_p = (8\pi GN)^{-1/2} =2.4 \times 10^{18}$ GeV in $D=4$. Here, we use a caret to indicate quantities  in the Einstein frame. One finds some transformed quantities that are useful in the calculation: 
\begin{align}
	\hat{g}^{\mu\nu} &= \Omega^{-2} g^{\mu\nu}, \quad \sqrt{- \hat{g}} = \Omega^D \sqrt{-g}, \nonumber \\ 
	\hat{R} &= \Omega^{-2} \left[ R - 2\frac{(D-1)}{\Omega} \square \Omega - \frac{(D-1)(D-4)}{\Omega^2} g^{\mu \nu} \nabla_\mu \Omega \nabla_\nu \Omega\right]
\end{align}
where $\square \Omega:= g^{\mu\nu}\nabla_\mu \nabla_\nu\Omega =\frac{1}{\sqrt{-g}}\partial_{\mu} \left[ \sqrt{-g} g^{\mu\nu} \partial_\nu \Omega\right]$. It is possible to get rid of the non-minimal coupling remaining  in action (\ref{4}) if we identify 
\begin{align}\label{omega}
	\Omega^{D-2} = \frac{2}{M_p^{D-2} }F_\chi. 
\end{align}
As a result, the action in the Einstein frame is

\begin{align}\label{5}
	\mathcal{S}_E = \int d^D x \sqrt{-\hat{g}} \left[ \frac{M_p^{D-2}}{2} \hat{R} -\frac{M_p^{D-2}}{2}\frac{(D-1)}{(D-2)}\frac{\hat{g}^{\mu\nu}}{F_\chi^2}  \hat{\nabla}_\mu F_\chi \hat{\nabla}_\nu F_\chi   - \frac{M_p^{D-2}}{2} \frac{ \hat{g}^{\mu \nu}}{F_\chi}  \hat{\nabla}_\mu \phi \hat{\nabla}_\nu \phi - V_E\right], 
\end{align}
where we have introduced the transformed potential  from the original one
\begin{align}\label{pot}
	V_E := \frac{M_p^{D}}{2^{D/(D-2)}} \frac{\left[V(\phi) - (F - \chi F_\chi)\right]}{F_\chi^{D/(D-2)}}. 
\end{align}
The transformed action (\ref{5}) still contains a non-canonical kinetic term in the second term of the equation and a mixing term to the scalar field $\phi$'s kinetic term. The former  can be further simplified to the canonical kinetic term by doing the field redefinition
\begin{align}
	\psi:= \sqrt{\frac{(D-1)}{(D-2)}} \ln\left( F_\chi \right), \quad \tilde{\psi}:= \sqrt{\frac{D-2}{D-1}} \psi.
\end{align}
Hence, after some treatments, we obtain the final action:   
\begin{align}
	\mathcal{S}_E = \int d^D x \sqrt{-\hat{g}} \left[ \frac{M_p^{D-2}}{2} \hat{R} -\frac{M_p^{D-2}}{2}\hat{g}^{\mu\nu} \hat{\nabla}_\mu \psi \hat{\nabla}_\nu \psi  - \frac{M_p^{D-2}}{2}  e^{ - \tilde{\psi}} \hat{g}^{\mu \nu}  \hat{\nabla}_\mu \phi \hat{\nabla}_\nu \phi - V_E\right]. 
\end{align}

\section{Starobinsky model with non-minimal coupling to a scalar field} 

In this section, we will investigate a particular example of $f(R)$ gravity theory that contains $R^2$ term called Starobinsky model \cite{Sta}. If we consider a nonminimally coupling to a scalar field, the gravitational Lagrangian density  can be written as $f(R) = \alpha(\phi) R + \beta(\phi) R^2$. Thus, we can find the associated Lagrangian density in term of an auxiliary field  $\chi$ as $F(\chi,\phi) = \alpha(\phi) \chi + \beta(\phi) \chi^2$ and we have  $\psi = \sqrt{3/2} \ln\left[\alpha + 2\beta \chi\right]$. The potential in the Einstein frame from equation  (\ref{pot}) take this following form
\begin{align}\label{pot2}
	V_E := \left(\frac{M_p^2}{2} \right)^2 \left[ e^{-\tilde{2\psi}}V(\phi) - \frac{\left(1-\alpha e^{-\tilde{\psi}}\right)^2}{2\beta}\right].
\end{align}
Therefore, there is a flat direction along  $\tilde{\psi} = \ln(\alpha)$ that the inflaton proceeds it on. Along this direction, the potential is given as
\begin{align}
	V = \frac{M_p^4}{4}  \frac{V(\phi)}{\alpha(\phi)^2}
\end{align}
showing that the inflation is effectively generated by a single field $\phi$ thus formulation corresponding to single field inflation can be used.

For a specific chase, we consider a scalar field with a general potential form $V(\phi) =  \mu_1\phi^2 + \mu_2\phi^4$ where $\mu_1 < 0$ to guarantee the field having a vacuum expectation value (VeV) at  $v:= \sqrt{-\mu_1/2\mu_2}$. In this model, it is also assumed a condition for non-minimal coupling to gravity to be  $\alpha(\phi) = a \phi^2 $. The potential (\ref{pot2}), therefore, can be written to be
\begin{align}\label{Vinf}
	V = V_0 \left[ 1- \mu_3 \tilde{\phi}^2\right]
\end{align}
where here we have defined $V_0:= \frac{\mu_2 M_p^4}{4 a^2} $, $\mu_3 := \frac{|\mu_1|}{\mu_2 M_p^2}$ and $\tilde{\phi} := \frac{M_p}{\phi}$. This is a kind of inverse power law inflation that is promising to explain  Planck experimental data for CMB and having a similar features as the  mutated hybrid inflationary models \cite{Zhun Lu}. Slow-roll parameters for this  single field inflation are:
\begin{align}
	\varepsilon = \frac{M_p^2}{2} \left[\frac{V'}{V}\right]^2 = \frac{2 \mu_3^2 \tilde{\phi}^6}{(1-\mu_3 \tilde{\phi}^2)^2},
\end{align}
\begin{align}
	\eta = M_p^2 \left[\frac{V''}{V}\right] = - \frac{6\mu_3 \tilde{\phi}^4}{(1-\mu_3 \tilde{\phi}^2)},
\end{align}
where  $V' := dV/d\phi$ and $V'' := d^2 V/d\phi^2$. Those slow-roll parameters show that the inflation starts, as indicated by condition $\varepsilon \ll 1$, at scale   $(\mu_3 \tilde{\phi}^2) \ll 1$ and terminates while $\varepsilon \simeq 1$ or at the scale  $(\mu_3 \tilde{\phi}_e^2) \simeq 1 + \sqrt{2}$. Observational parameters, the scalar spectrum index $n_s$ and the ratio of the tensor perturbation to the scalar perturbation $r$, can be represented by using those two slow-roll parameters as follows \cite{slowroll}
\begin{align}
	n_s = 1-6\varepsilon + 2 \eta, \quad r=16\varepsilon
\end{align}

During that stage, the inflaton evolves from a particular value $\phi$ into  its value at the end of inflation  $\phi_e$ as given by  e-folding number 
\begin{align}
	N &= -\frac{1}{M_p^2} \int^{\phi_e}_{\phi} \frac{V}{V'} d\phi = \frac{1}{2\mu_3}\left[ \frac{\tilde{\phi}^{-4}}{4} - \frac{ \mu_3\tilde{\phi}^{-2}}{2}\right]_{\tilde{\phi}_e}^{\tilde{\phi}} \nonumber \\
	&= N(\tilde{\phi}) - N(\tilde{\phi}_e).
\end{align}
where here, we have defined 
 \begin{align}
	 N(\tilde{\phi}) := \frac{1}{2\mu_3}\left[ \frac{\tilde{\phi}^{-4}}{4} - \frac{ \mu_3\tilde{\phi}^{-2}}{2}\right].
 \end{align}

The spectrum of scalar perturbation predicted by the inflation 
is expressed as \cite{slowroll}
\begin{align}
\mathcal {P}_\mathcal{R}(k)=\Delta_{\mathcal{R}}^2\left(\frac{k}{k^\ast}\right)^{n_s-1},  \qquad
\Delta_{\mathcal{R}}^2=\frac{V}{24\pi^2M_{\rm pl}^4\varepsilon}\Big|_{k^\ast}. 
\label{power}
\end{align}
The CMB observations give the normalization such that 
$\Delta_{\mathcal{R}}^2\simeq 2.43\times 10^{-9}$ at $k_\ast=0.002~{\rm Mpc}^{-1}$. 
This constrains the value of $V/\varepsilon$ at the time when 
the scale characterized by the wave number $k_\ast$ exits the horizon \cite{uobs}.
On the other hand, the remaining e-foldings $N_\ast$ of the inflation 
after the scale $k_\ast$ exits the horizon is dependent on the reheating 
phenomena and others as \cite{slowroll, Rub}
\begin{equation}
N_\ast\simeq 61.4-\ln\frac{k_\ast}{a_0H_0}
-\ln\frac{10^{16}~{\rm GeV}}{V_{k_\ast}^{1/4}}
+\ln\frac{V_{k_\ast}^{1/4}}{V_{\rm end}^{1/4}}-
\frac{1}{3}\ln\frac{V_{\rm end}^{1/4}}{\rho_{\rm reh}^{1/4}}.
\end{equation} 
Taking account of this uncertainty, $N_\ast$ is usually considered to 
take a value in the range 50 - 60.  Here we also use the values in this range and we represent a value of $\phi$ which gives the e-foldings $N_\ast$ as $\phi_\ast$. The observational observables, tensor to scalar ratio $r$ and the spectral index $n_s$, need to be expressed at this value to be able have comparison with the observation. These values will be denoted as $r_\ast $ and $n_{s\ast}$. 

\begin{figure}[htp]
\begin{center}
\includegraphics[width=0.80\textwidth]{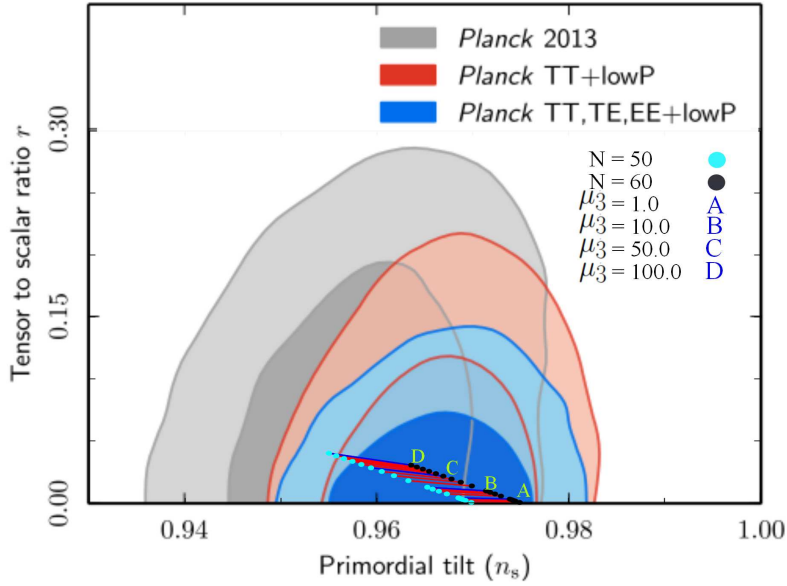}
\end{center}
\vspace*{5mm}
\caption{\label{Gb} Predicted values of ($n_s, r$) for various  $\mu_3$ values in the interval $(0, 100)$  as given in Table 1. Contours given as Fig. 6 in Planck Collaboration
 \cite{Planck2015} are used here.}
\end{figure}

\section{Predictions}

To discuss the features of the model, numerical calculations are needed particularly to consider all possible terms in the formula. However, in this model,  the inflation  ends at scale  $ \tilde{\phi}_e\simeq \sqrt{(1 + \sqrt{2})/\mu_3}$. Therefore, for $\mu_3 < 10^2$, $N_e = N(\tilde{\phi}_e)$ is negligible thus e-folding number can be approximately given as it value when the comoving scale crosses the Hubble radius at the first time $N\simeq N_\ast$.  Considering that the inflation starts at the scale $(\mu_3 \tilde{\phi}^2) \ll 1$, the leading contributions of the e-folding number and slow-roll parameters are 
\begin{align}
	N_\ast &\simeq \frac{\tilde{\phi}^{-4}}{8 \mu_3}, \\
	\varepsilon_\ast &\simeq 2 \mu_3^2 \tilde{\phi}^6 = \sqrt{\frac{\mu_3}{2^{7}} }\frac{1}{N_\ast^{3/2}}, \\
	\eta_\ast &\simeq - 6 \mu_3 \tilde{\phi}^4 = - \frac{3}{4 }\frac{1}{N_\ast}.
\end{align}
 The scalar-to-tensor ratio $r_\ast$ and the spectral index $n_{s\ast}$ therefore will depend to $N_\ast$ and $\mu_3$ as follows
 \begin{align}
	 n_{s\ast} = 1-6\sqrt{\frac{\mu_3}{2^{7}} }\frac{1}{N_\ast^{3/2}} -  \frac{3}{2 }\frac{1}{N_\ast}, \quad
	r_\ast =16 \sqrt{\frac{\mu_3}{2^{7}} }\frac{1}{N_\ast^{3/2}}.
 \end{align}

CMB maps provided by  Planck 2015 constrains the tensor-to-scalar ratio and the spectral index  given as $n_s = 0.968 \pm 0.006$ and $r< 0.11$ (95$\%$ CL) respectfully \cite{Planck2015}. Due to these constraints, several numerical results are shown in the table 1~ for some value of $\mu_3$ with a fixed $N_\ast = 50, 60$, as well as in the Figure \ref{Gb} for the predicted points $(n_s, r)$ in the contour provided from Planck 2015.   

\begin{figure}[t]
\begin{center}
\begin{tabular}{|c|c|c|c|c|c|}\hline
\hline

$ \mu_3$ & $N_\ast$ &$n_{s\ast}$ & $r_\ast$ &VeV ($v$)&$a$\\
           &          &            &           &  $\times M_p$    & $\times 10^{4}$ \\
\hline
\hline
0.01& 50.0 &    0.96985  & 0.00040  & 0.07070 &13.1755  \\
0.01& 60.0 &    0.97489  & 0.00030  & 0.07070 &15.1061  \\
\hline
1.00& 50.0 &    0.96850  & 0.00400  & 0.70711 &4.16645   \\
1.00& 60.0 &    0.97386  & 0.00304  & 0.70711 &4.77697  \\ 
\hline
10.0& 50.0 &    0.96526  & 0.01265  & 2.23607 &2.34297 \\   
10.0& 60.0 &    0.97139  & 0.00962  & 2.23607 &2.68629 \\   
\hline  
30.0& 50.0 &    0.96178  & 0.02191  & 3.87298 &1.78027 \\    
30.0& 60.0 &    0.96875  & 0.01667  & 3.87298 &2.04114 \\   
\hline   
50.0& 50.0 &    0.95939  & 0.02828  & 5.00000 &1.56684 \\   
50.0& 60.0 &    0.96693  & 0.02152  & 5.00000 &1.79643 \\    
\hline
75.0& 50.0 &    0.95701  & 0.03464  & 6.12372 &1.41580 \\   
75.0& 60.0 &    0.96512  & 0.02635  & 6.12372 &1.62326 \\   
\hline  
100.0 & 50.0  &   0.95500  & 0.04000  & 7.07107 &1.31755 \\    
100.0 & 60.0  &   0.96359  & 0.03043  & 7.07107 &1.51061 \\    
\hline

\hline
\end{tabular} \\
\vspace*{5mm}
{Table~1\ Numerical results for several variations of $\mu_3$ and $N_\ast$. }
\end{center}
\end{figure}

The examples show that the predicted points match nicely in the $1\sigma$ confidence level of \textit{Plank} TT, TE, EE+lowP combination data for $ 0<\mu_3 \leq 100$, for which  the vacuum expectation value $v$ of the scalar field $\phi$ and non-minimal coupling constant to the $R$ gravity term $a$ can be estimated.  The latter one can be approximated from the CMB normalization given in  equation (\ref{power}). The scalar generating the inflation, therefore, should evolved in the transPlanckian value and having vacuum expectation value $v=\sqrt{\mu_3/2} M_p$. The non-minimal coupling is typically in $10^{4}$ order which is in the same order of non-minimal coupling of the Higgs inflation scenario \cite{Higgs}.  This large value had been shown causing several difficulties such as unitarity problem  \cite{Unitary2, Unitary3, Unitary1}, however some treatments can be used to cure the problem \cite{antiUn1, antiUn2}. The reason might be from the absence of the non-minimal coupling to $R^2$ gravity term when the  inflaton takes the flat direction in the potential (\ref{Vinf}), therefore using other scenario in this context may treat the problem. Instead of the difficulty that still appears, the main purpose to save scalar power law potentials which are already disfavored by recent observations can be achieved.



\begin{thebibliography}{9}

\bibitem{Planck2015} Planck Collaboration (P.A.R. Ade (Cardiff U.) et al.), \textit{Planck 2015 results. XX. Constraints on inflation}, Astron.Astrophys. 594 (2016) A20, arXiv:1502.02114 [astro-ph.CO].  

\bibitem{Higgs} Fedor L. Bezrukov and Mikhail Shaposhnikov, \textit{The Standard Model Higgs boson as the inflaton}, Phys.Lett. B659 (2008) 703-706, arXiv:0710.3755 [hep-th].	

\bibitem{Unitary2} Burgess, C.P. et al.	\textit{Power-counting and the Validity of the Classical Approximation During Inflation}, JHEP 0909 (2009) 103, arXiv:0902.4465 [hep-ph] 

\bibitem{Unitary3} Barbon, J.L.F. et al, \textit{On the Naturalness of Higgs Inflation}, Phys.Rev. D79 (2009) 081302, arXiv:0903.0355 [hep-ph]. 

\bibitem{fR1} Girish Chakravarty, Subhendra Mohanty, Naveen K. Singh, \textit{Higgs Inflation in $f(\phi,R)$ Theory}, Int. J. Mod. Phys. D 23, 1450029 (2014), arXiv:1303.3870 [astro-ph.CO].

\bibitem{Mod} G. Chakravarty, S. Mohanty and N. K. Singh, Int. J.
Mod. Phys. D 23, no. 4, 1450029 (2014); J. Joergensen,
F. Sannino and O. Svendsen, arXiv:1403.3289 [hep-ph];
A. Codello, J. Joergensen, F. Sannino and O. Svendsen,
arXiv:1404.3558 [hep-ph]; G. K. Chakravarty and S. Mohanty,
Phys. Lett. B 746, 242 (2015).

\bibitem{fR2} de la Cruz-Dombriz, Alvaro et al., \textit{Spotting deviations from $R^2$ inflation}, JCAP 1605 (2016) no.05, 060 arXiv:1603.05537 [gr-qc]. 

\bibitem{fR3} Torabian, Mahdi,\textit{When Higgs Meets Starobinsky in the Early Universe}, IPM-PA-381,  arXiv:1410.1744 [hep-ph].
 
\bibitem{fR4} Sho Kaneda and Sergei V. Ketov,  \textit{Starobinsky-like two-field inflation}, Eur. Phys. J. C (2016) 76:26.


\bibitem{Conf} Kaiser, David I, \textit{Conformal Transformations with Multiple Scalar Fields}, Phys.Rev. D81 (2010) 084044 arXiv:1003.1159 [gr-qc] MIT-CTP-4125

\bibitem {Sta} A. A. Starobinsky, Phys. Lett. B 91 (1980) 99.

\bibitem{Zhun Lu} Zhun Lu,  \textit{Inflation in the Generalized Inverse Power Law Scenario}, JCAP 1311 (2013) 038,  arXiv:1311.0348 [hep-th].

\bibitem{slowroll} D.~H.~Lyth and A.~Riotto, Phys. Rep. {\bf 314} (1999) 1; A.~R.~Liddle and D.~H.~Lyth, {\it Cosmological Inflation and Large-Scale Structure} (Cambridge, 2000).

\bibitem{Rub} D. S. Gorbunov, V. A. Rubakov, \textit{Introduction to the theory of the early universe, Cosmological perturbations and inflationary theory}, World Scientific, 2011.


\bibitem{uobs}WMAP Collaboration, D.~N.~Spergel, {\it et al.}, \emph{First year Wilkinson Microwave Anisotropy Probe (WMAP) observations: Determination of cosmological parameters}.

\bibitem{Unitary1} F. Bezrukov, A. Magnin, M. Shaposhnikov (ITPP, Lausanne), and S. Sibiryakov, \textit{Higgs inflation: consistency and generalisations}, JHEP 1101 (2011) 016, arXiv:1008.5157 [hep-ph].

\bibitem{antiUn1} Lerner, Rose N. et al., \textit{A Unitarity-Conserving Higgs Inflation Model,} Phys.Rev. D82 (2010) 103525, arXiv:1005.2978 [hep-ph].

\bibitem{antiUn2}Hyun Min Lee, \textit{Running inflation with unitary Higgs}, Phys.Lett. B722 (2013) 198-206, arXiv:1301.1787 [hep-ph]. 




\end{thebibliography}
\end{document}